# COMIT

## Cryptographically-secure Off-chain Multi-asset Instant Transaction network

Making global payments as cheap, fast and easy as sending a text message.

**Abstract:**

Since the creation of Bitcoin in 2009 we have seen a great push towards public and private blockchains. In order to avoid fragmentation, a global network connecting all these blockchains is envisioned. Just like the Internet facilitates communication and the transfer of information, we propose a system, similar in size and reach for payments and transactions: A cryptographically-secure off-chain multi-asset instant transaction network (COMIT) can connect and exchange any asset on any blockchain to any other blockchain using a cross-chain routing protocol (CRP).

COMIT is a super blockchain network that allows for instant transactions which are enforced using off-chain smart contracts. It leverages Payment Channels and Hashed Timelock Contracts (HTLC) across chains to solve the problem of double spending attacks without requiring a settlement onto the underlying blockchains. COMIT's connectivity is provided by Liquidity Providers (LP), who operate on one or more blockchains, acting as payment hubs and nodes on a single chain and market makers in a decentralized network for cross-chain asset conversions.

This paper lays out how COMIT works, the benefits for Users, Liquidity Providers and Businesses; and how this does not only accelerate the adoption of blockchain technology, but furthermore allows for an integration with the traditional banking system.

# 1. Introduction

In this paper we are going to discuss the development of the Internet of transactions also called COMIT (Cryptographically secure Off-chain Multi asset Instant Transaction network). Before we describe a general overview and technical details, we are going to layout the current financial and economic challenges, the banking and blockchain world are facing today. In order to be on the same page, we will start with the status quo of both of these areas.

## 1.1. Ecosystem overview

### 1.1.1. Blockchain Ecosystem

Since the start of Bitcoin[1] in January 2009, we have seen the introduction of a multitude of blockchains across all kinds of areas and financial markets. Today we can count hundreds of public blockchains that amount to a total market cap of around 20 Billion dollars, excluding many more private blockchain installations.

Last year we saw the emergence of precious metal backed tokens, derivatives, entirely new asset classes representing entire ecosystems, and even ETF tokens to invest into other blockchain assets. One such example are Initial Coin Offerings (ICOs) that are gaining in popularity. The World Economic Forum is even going as far as predicting that 10% of the global GDP will be stored on the blockchain in less than 10 years[2]. In terms of today's global GDP that would be $7.8 trillion.

One of the other big promises of blockchain technology today, is the complete automation of financial business logic via smart contracts. The Ethereum[3] blockchain and its ecosystem is noteworthy for their "Turing-complete" blockchain, which allows any smart contract to be executed on-chain, thereby allowing the trustless execution of any business logic.

Bitcoin on the other hand has a very limited smart contracting language and primarily uses cryptographically secure off-chain smart contracts for such logic. A great example of such off-chain smart contracts is the Lightning Network[4], [4] which is seeing an increase in popularity as a possible solution to the Bitcoin scaling problem.

### 1.1.2. Bank Ecosystem

Banks have seen increasing challenges with the ever changing financial environment. Some traditional players and some newcomers have already adopted new technologies to cope with these new challenges. Banks like N26[5] in Europe or Simple[6] in the US are using low cost online banking infrastructure that is more efficient and easier to use for its users. Traditional banks such as DBS, who was awarded World's Best Digital Bank[7], and many others are adapting their business models to better cater to their customers' needs by bringing many of their services onto mobile platforms and the Internet.

Another solution banks have come up with to adapt sufficiently is the introduction of their own private blockchains and distributed ledger technologies (DLT). Instead of participating in the public blockchains like Bitcoin and Ethereum, they choose to run the technology in-house to retain full control over their business logic.

## 1.2. Challenges of the current system of financial transactions

The above mentioned technologies and inventions are being introduced to better cater to end-users (U), bring more advantages to banks (and other Liquidity Providers (LP)) and businesses (B). We would like to list a few in regards to each of these three parties.

1.2.1. Users [U]
- are looking for the cheapest possible way to execute their financial transactions. Be it a bank transfer to their beloved ones, or buying stocks in their favorite companies. To save costs, banks are therefore onboarding their users onto a cheaper online banking environment. At the same time, the costs of cryptocurrency transactions have been reduced greatly - to a fraction of traditional banking fees.
- are also looking for the fastest execution of a financial transaction. Central banks in many countries have deployed real-time payment networks, which are vastly faster compared to traditional bank transfers that used to take multiple days to complete. Today's public blockchains either require the a user to wait a couple of minutes or a maximum of a few hours; bringing a great advantage to the speed of global transactions.

1.2.2. Liquidity Providers [LP] (defined in detail the next chapter)
- are starting to understand the challenges new technologies bring to their long-standing business model. Private blockchains seem to offer this long awaited alternative.
- are incurring massive costs of human capital. A change to digital payment gateways is a welcome relief.
- are facing more and more competition in their so far well-protected market segment. Whoever is open minded into adopting new business models will keep riding the wave of success. Private Blockchain and Distributed Ledger Technologies appear promising in that regard.

1.2.3. Businesses [B]
- are looking to increase revenue and customer adoption. Globalization and new technologies seem to provide solutions to both these needs.

## 1.3. Solutions with infrastructural problems

Riding a horse seemed to be a great advantage to only being on foot, but then the automobile came, and mass-manufacturing and made horses obsolete. The person who bought 500 horses the day the automobile started to be mass-manufactured by Ford, lost greatly. Exactly in the same way, some of the aforementioned attempts to solve the challenges users, banks and businesses are experiences, are more like a trojan horse with an underlying infrastructural problem, rather than an actual solution. Let us elaborate.

The emergence of online banking has been a step forward in the right direction. However speed and cost will once again become a limiting factor. Therefore more and more users will appreciate the up-sides of speed and trust that a decentralized blockchain system provides. With that in mind banks and many other groups start to create individual private blockchains, which is then leading to a fragmented environment instead of mutual collaboration.

Public blockchains themselves have problems as well: Transaction costs, even though being much lower than bank transfers, are increasing with growing usage. Payments that require instant confirmation such as checking out a high ticket item in a store can not work via the traditional blockchain, as the current required confirmation time of even just a few minutes is not feasible in a point of sales (POS) environment, where speed is critical.

For users, a couple of challenges are arising. They like the up-sides of the online-banking system, however, they increasingly realize that the future lies in the decentralized blockchain environment. Yet adoption rates of these new systems remain relatively low, mostly because these users need to trust new parties (exchanges, wallets and payment providers) who many times do not offer user-friendly solutions. Banks feel being left out and can not support the adoption process either.

This moves the intended easy global access further apart, rather than closer together. The dream of financial inclusion of low-income countries into the global system of transactions is thereby drifting further and further away. This is bad for users, banks and businesses the same.

What is going to be the inevitable solution:

# 2. COMIT - the Cryptographically secure Off-chain Multi-asset Instant Transaction network

Before we go into the details of what COMIT is, why and how it works, we would like to discuss a similar model that was adopted over the past 30 years:

## 2.1. Comparing the Internet to COMIT

2.1.1. The Internet - the Network of Information and Communication

Before the invention of the TCP/IP protocol the Internet was dispersed in many local networks, so-called Intranets. These provided local efficiency over the more traditional point-to-point communication (such as letter, fax, telephone calls). The real breakthrough only came in 1973, when different Intranet networks realized that they could use a unifying Internetwork protocol to communicate among each other, thereby extending reach by compatibility even more.

With the requirements for an Intranet to join the so called Internet dropping to the bare minimum, it became possible to add almost any Intranet, no matter how basic or sophisticated their characteristics were.

The initial adoption by users was relatively slow, as the services offered at the beginning were limited. There was one major factor however, that eventually sped it up significantly. The same providers that were already offering mail, FAX and phone services, could now add Internet services to their portfolio giving them extra revenue streams. User adoption came easily, as a trust basis between the customers and these services providers was already established for years or even decades. Early adopters started, the late adopters followed.

Today the Internet spans across the entire world and information that used to be accessible only locally is now accessible from anywhere, even from the moon . Information is stored by servers all over the world while routers create the backbone. Internet service providers (ISP) give the average end-user easy and quick access to this vast database of information by opening a communication channel to their customers and to other ISPs, servers and routers.

Once the average user accesses the Internet through his or her communication channel with the ISP in order to gain information from the Internet, the user does not have to worry about how the information is retrieved exactly. All she has to do, is to type in the destination from where she wants to retrieve the information (URL). The ISP, to which she has the communication channel to, does not know the exact path to the destination either. However, through the TCP/IP protocol, the request is routed through from one communication channel to another using routers, servers or ISPs, who then either know the location or continue the process. The important point is, neither one of them has to know the entire way. All they have to do, is to trust the TCP/IP protocol, which has the task of delivering packets from the source host to the destination host, solely based on the IP addresses in the packet headers. Its routing function enables internetworking, and essentially establishes the Internet[8].

How does this translate into the Internet of transactions?

2.1.2. COMIT - the Internet of Transactions

The basic structure of the Internet and COMIT is exactly the same as their purposes are similar; the exchange of something. In today's world, the exchange of value works similar as the exchange of information pre-Internet; point-to-point in an enclosed system. As described in the beginning we have archaic banking systems that do not allow for easy access or transfers from one asset to another. Including all the other additional challenges discussed above. Many of the suggested solutions we have recently seen, do not provide the same final and elegant solution as the Internet did for information.

With that in mind, we therefore suggest a Cryptographically-secure Off-chain Multi-asset Instant Transaction network (COMIT). What does such a network look like? Just like in the Internet, we need a stable and trustworthy backbone. In our opinion any large blockchain provides exactly that. It can be any blockchain, because just like on the Internet, different modalities will be interconnected (For example: the initial Internet never foresaw mobile app messaging services, but these have been implemented without any problems). The same will be true for COMIT, where any new blockchain can be connected to an existing one through the use of the COMIT Routing Protocol (CRP).

A user today, who is using crypto-currencies, currently has to wait minutes if not hours before a transaction is accepted by the counterparty. With the adoption of payment channels, such as the Lightning Network, Raiden or many others, such users can transfer assets instantly from person A to person B. If person B then opens another payment channel to person C, person A can also transfer assets to person C via B instantaneously, as long as person B provides enough liquidity. In theory there can be an infinite chain of participants in between person A and C, as long as they all provide enough liquidity. Again, such transactions are immediate without person A needing to know which route the assets took to end up at person C. She can trust this system as the routing protocol ensures its correctness, plus the cryptographically secured payment channels, which will be described in the next chapter, ensures flawless functionality.

What we end up with, are cryptographically-secured instant payments off-blockchain that can even be transferred from one asset to another via hashed time-lock contracts (those will also be described in the next chapter). In order for this network to have enough liquidity (in the example above person B needs to provide enough liquidity to enable a transaction between person A and person C), we introduce the concept of Liquidity Providers (LP). LPs can be seen or understood as hubs or nodes in the COMIT network, that create payment channels to users, other LPs and businesses. They are a core part in COMIT. Just like servers, routers and ISPs are to the internet.

So how does the big picture fit together? In today's traditional blockchain ecosystems, banks are often left out, in COMIT on the contrary, banks, exchanges and many others can take over the role of Liquidity Providers = ISPs on the Internet. Instead of users having to learn and trust new systems or companies, just like in the early adoption phase of the internet, they can rely on a partner that they are already comfortable with and trust. Adoption of this system will be seamless, fast and will bring great benefits to all of its participants, just like the Internet did. Some of the benefits of COMIT include:, but are not limited to:
- Open source infrastructure
- True instant, frictionless and cheap payments for users all over the world
- True global access without limitations to any asset or business process connected to a blockchain
- Cryptographically secure trustless global transactions network
- Amazing new business opportunities for companies
- New recurring revenue streams for banks and other liquidity providers
- Rapid adoption based on existing networks build with new cheap and secure infrastructure

Our vision for the world is as follows: Sending money will be as cheap and seamless as sending a WhatsApp message.

## 2.2. Technical Details

This chapter is for the technical reader who is well versed with how blockchains work. For illustration purposes, we will use Bitcoin style transactions, which require a full spend of every output. Account based blockchains, can be connected in COMIT as well, but require a slightly different form of payment channel (See Raiden Network[9] for details).

The following details the minimum requirements for a blockchain to be compatible with COMIT:

2.2.1. Minimal Requirements for a blockchain to be COMIT-compatible

The basic requirements for COMIT-compatibility are two-fold:
1. Routing related: a basic hash function, as well as time-locks are required to route across different chains.
2. Speed & cost related: payment channel support is required to reduce cost & make transactions instant. Federated sidechains & private chains which already offer instant transactions, can be considered to skip this requirement.

2.2.1.1. Double-Spend Protection

Double-spend protection is the main reason blockchains exist in the first place. In technical terms this means that two valid transactions which spend the same transaction output (UTXO), will conflict and only one can be confirmed in the network. Account based blockchains (for example Ethereum) that allow for spending the same amount from the same address multiple times, usually have other means to prevent double spending.

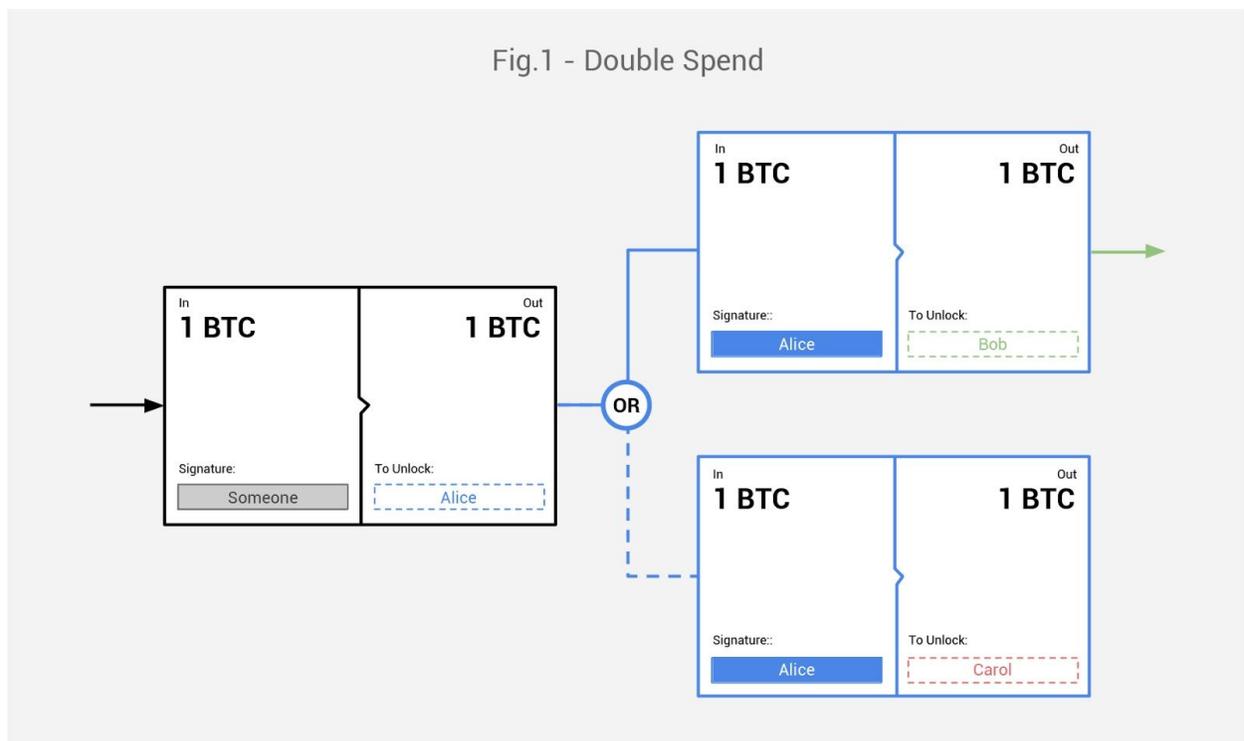

### 2.2.1.2. Multi-signature (multisig)

Multisig is a very old and well-trusted concept that can be compared to a shared checkbook with multiple required signatories. A multisig transaction allows to enforce arbitrary joint signature rules. COMIT uses 2 out of 2 multisig transactions for which both signers have to sign a transaction to become valid and be accepted by the network. Multi-signature transactions are a requirement for Payment Channels.

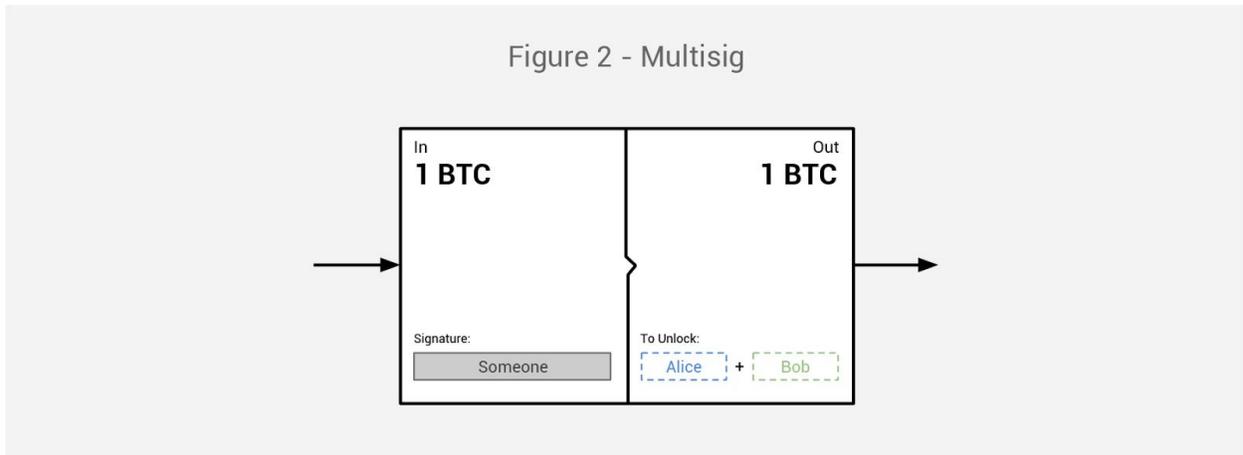

### 2.2.1.3. Time-Locks

A timelock is a simple requirement for funds to be locked up until a future date. Blockchains are found to have 2 different kind of time-locks: relative and absolute time-locks. Absolute time-locks will lock a transaction output until a fixed time in the future. Relative time-locks will lock a transaction output relative to the time the transaction was confirmed. Time-locks are a requirement for trustless Payment Channels and relative time-locks are recommended as they allow for indefinitely open Payment Channels.

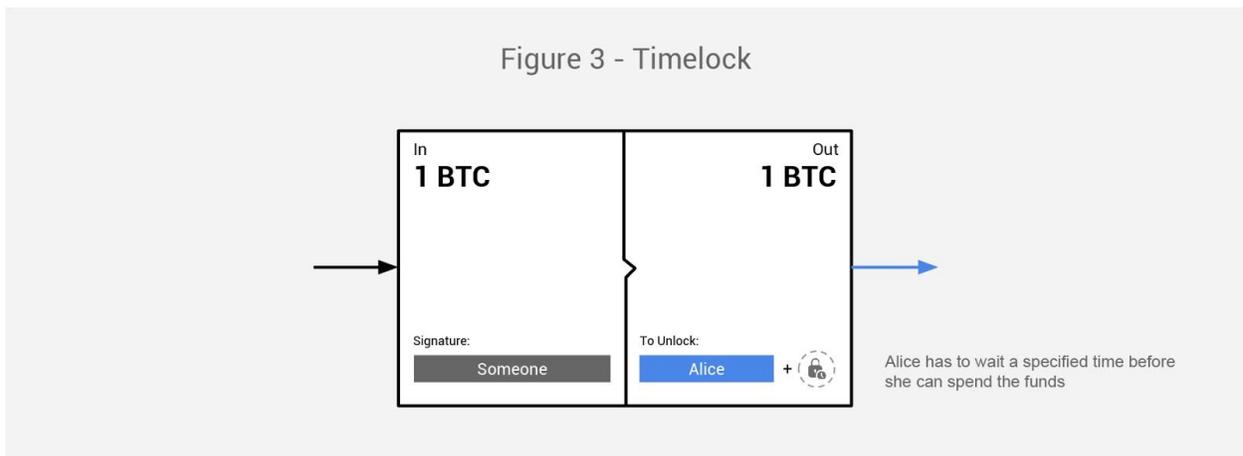

### 2.2.1.4. Hash function

Hash functions are a standard cryptographic concept. They are one-way functions to convert arbitrary data (in our case a *secret s*) into a *hash h*. The *hash h* can then be shared safely without anyone being able to compute the *secret s* used to create it. This allows us to build a hash-lock transaction which will only unlock funds with the knowledge of the *secret s*.

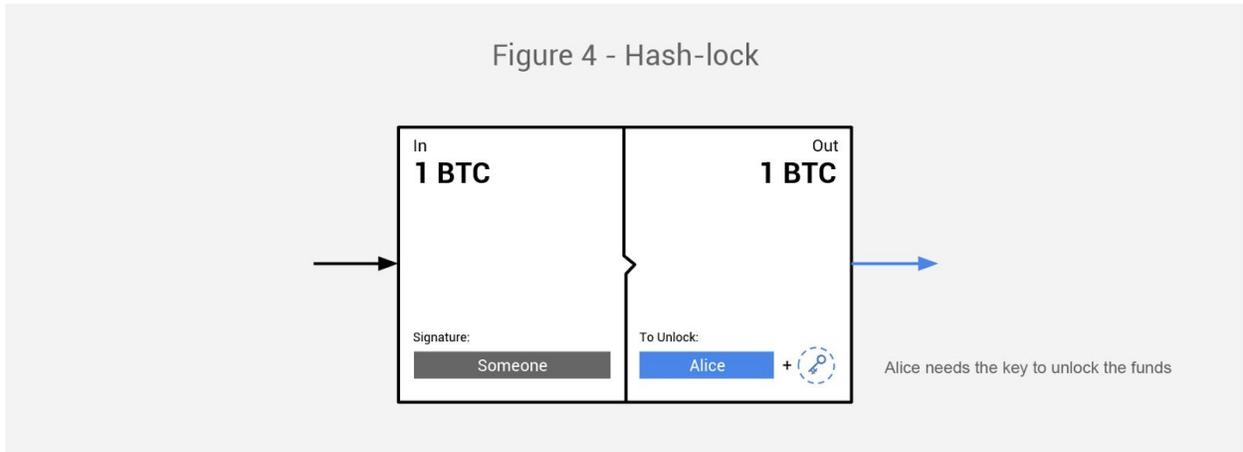

In order to route across multiple blockchains, we need the same hash function available in the smart contracting language of each blockchain participating on such a route.

### 2.2.2. Building Blocks for COMIT

#### 2.2.2.1. Hashed Time-lock Contracts

A hashed time-lock contract (HTLC[4]) combines the concept of a time-lock for refund purposes with a hash-lock. If the recipient can provide the *secret s* for the hash lock before the expiry of the time-lock, he will be able to retrieve the funds. Otherwise the sender can safely reclaim the funds.

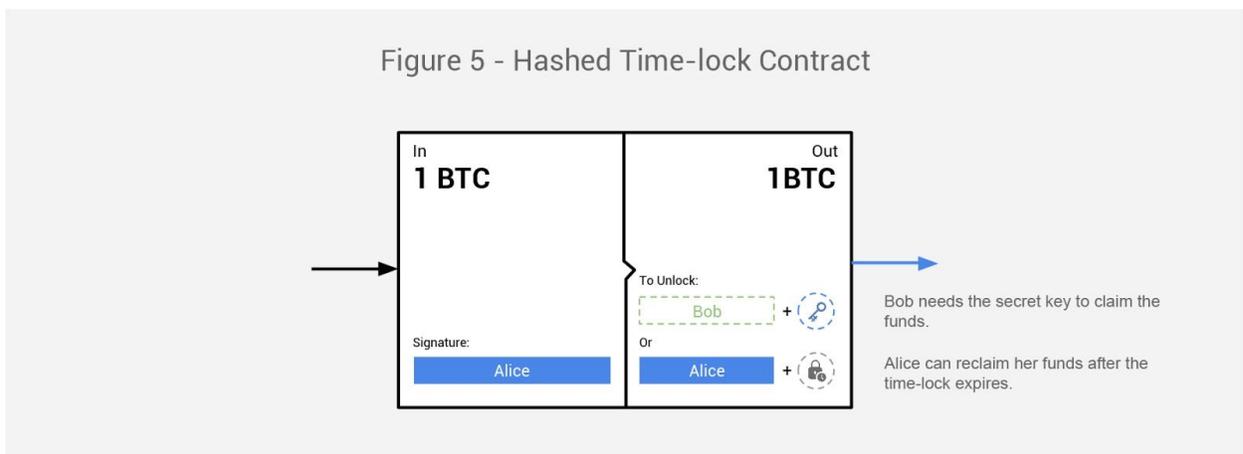

This HTLC can then be used to link two payment channels together. The link mechanism is the same *secret s*, which is initially created by the recipient. Subsequently the receiver will share the hash of *s* with the sender who will create a conditional transaction, which has an output that can be redeemed with the knowledge of the *secret s*. Every node in the Payment Channel chain can then safely use the same hash to create a transaction which is also conditional on knowing the *secret s*. In the end you have a chain of transactions which all depend on the same secret to be full-filled. When the receiver takes the last transaction and uses the secret to redeem the money, every other node will see the secret that was used and can then fulfill their own incoming transaction.

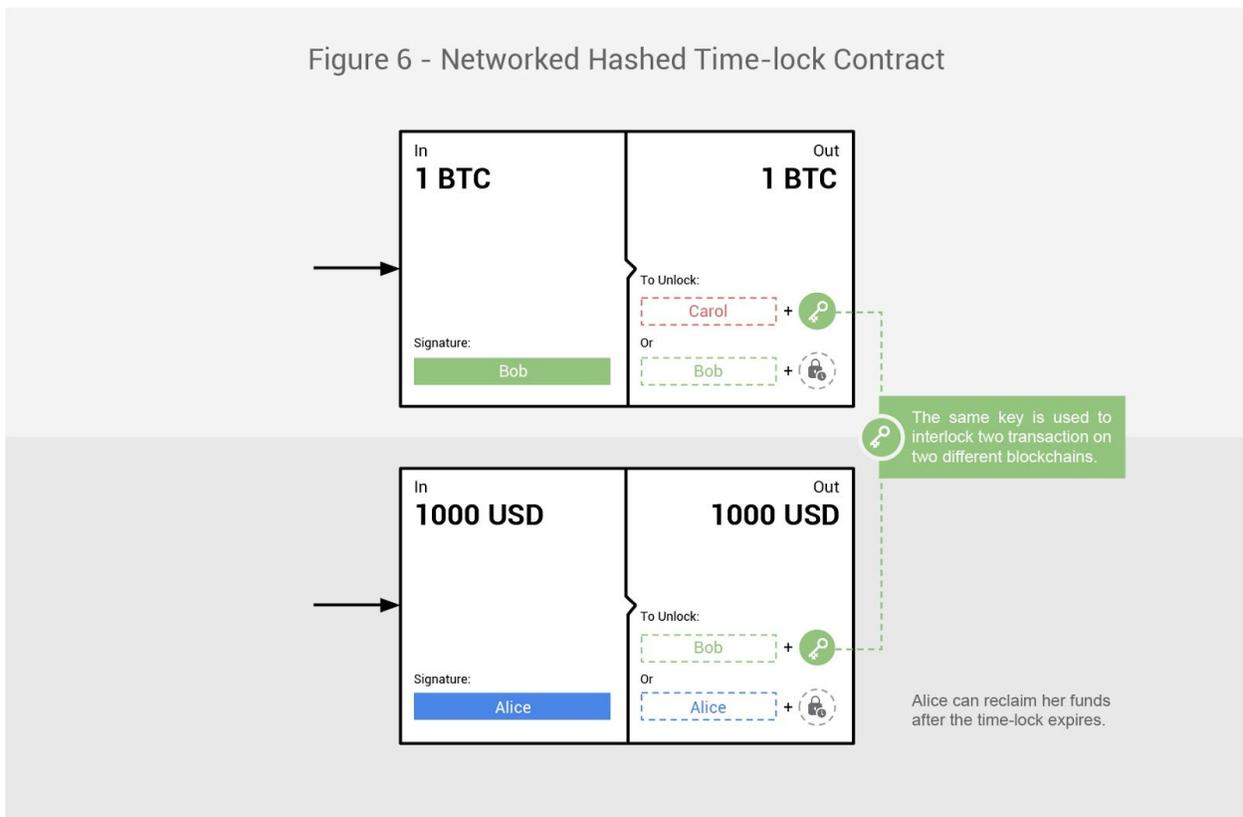

Figure 6 - Networked Hashed Time-lock Contract

The time-lock mechanism is used as a refund mechanism in case of an intermittent routing failure. The time-locks need to be stacked from receiver towards sender to make sure no-one is able to cheat.

2.2.2.2. Payment Channels

Payment Channels are the basic building blocks of the COMIT network. As previously discussed, Payment Channels can take different forms depending on the blockchain being used. For illustration purposes we use the UTXO model of the Bitcoin blockchain.

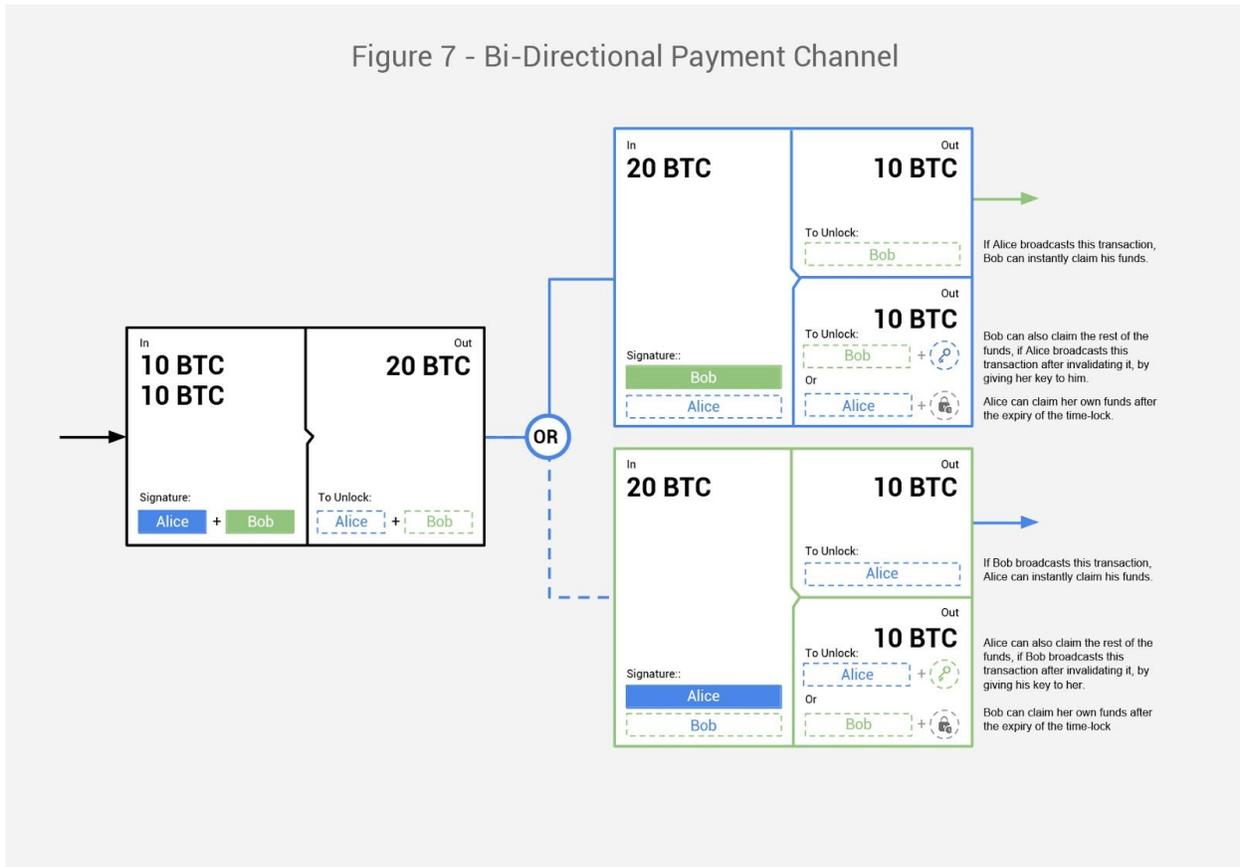

Figure 7 - Bi-Directional Payment Channel

2.2.3. Cross-chain Payment Channels

COMIT uses HTLCs and Payment Channels for cross-chain transactions. LPs place bids for asset exchange across channels and the COMIT Routing Protocol (CRP) selects the best path across multiple LPs to send the transaction to the recipient.

A cross-chain transaction therefore involves three steps:

2.2.3.1. Initial route finding using the CRP (see next chapter for details)

Routes can transfer across many different LPs and go across multiple different blockchains. Such a route includes the hashing capabilities of the blockchains used and needs to use the same hash function across all Payment Channels used for this transaction.

### 2.2.3.2. Sending the transaction across chains using HTLCs

After the route has been determined, all the participating Payment Channels are connected via HTLCs. To do this, the recipient creates a *secret s* and hashes it using the selected hash function. This hash is initially shared with the sender, who will then subsequently send a conditional payment to the first LP requiring knowledge of the secret s to redeem. Each LP in the route can then safely forward the transaction while adding the same conditional to the transaction redemption. Through the use of HTLCs we can guarantee that either the entire transaction via this route gets fulfilled or all Payment Channel transactions will be unredeemable. No trust has to be put in any of the LPs in the middle of the route. As a last step, upon receiving the conditional payment from the last LP in the chain, the receiver shares the *secret s* with the sender and all the LPs involved in the transaction. This is the final acknowledgement of the transaction receipt.

### 2.2.3.3. Settling the HTLC transactions into the underlying channels

After the *secret s* has been shared across the route, every Payment Channel will then settle the transaction back into the channel. This is done by updating the Payment Channel state to the final balances and then invalidating the HTLC transaction by revealing the *invalidation key k* to the Payment Channel counterparty.

# Cross Chain Bi-Directional Payment Channel with HTLC

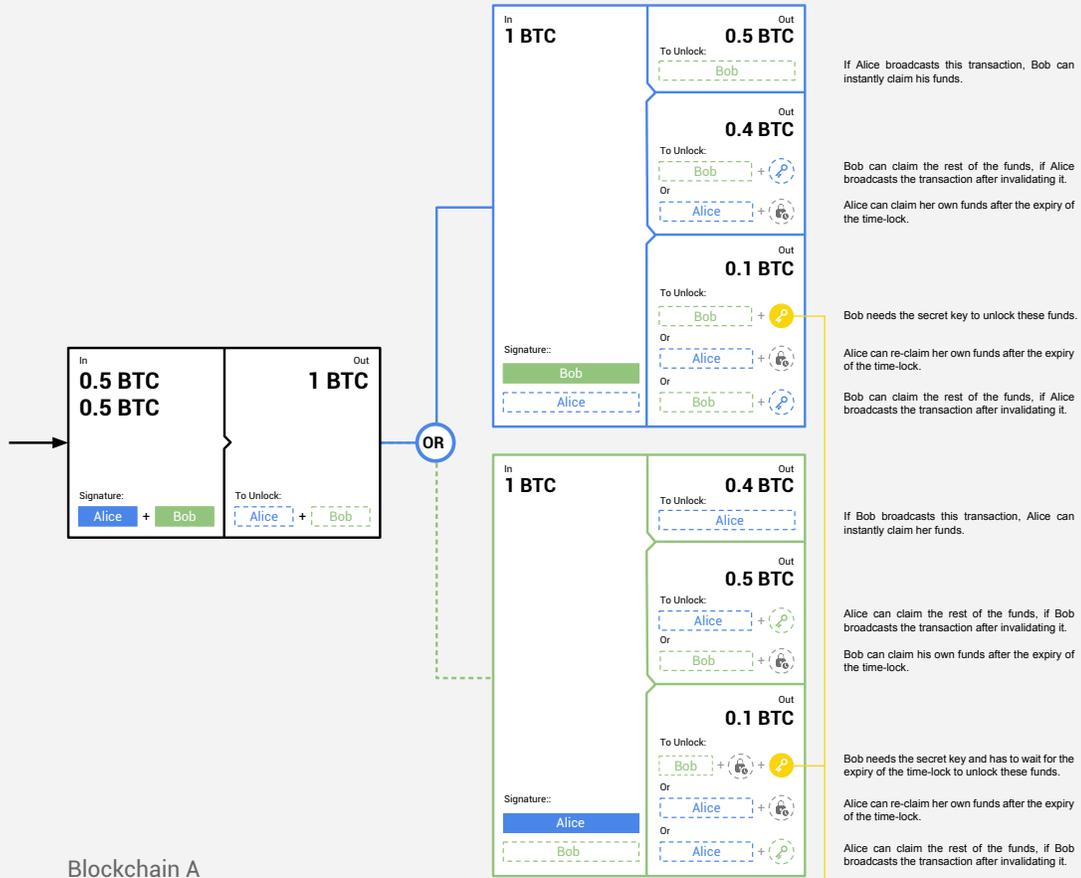

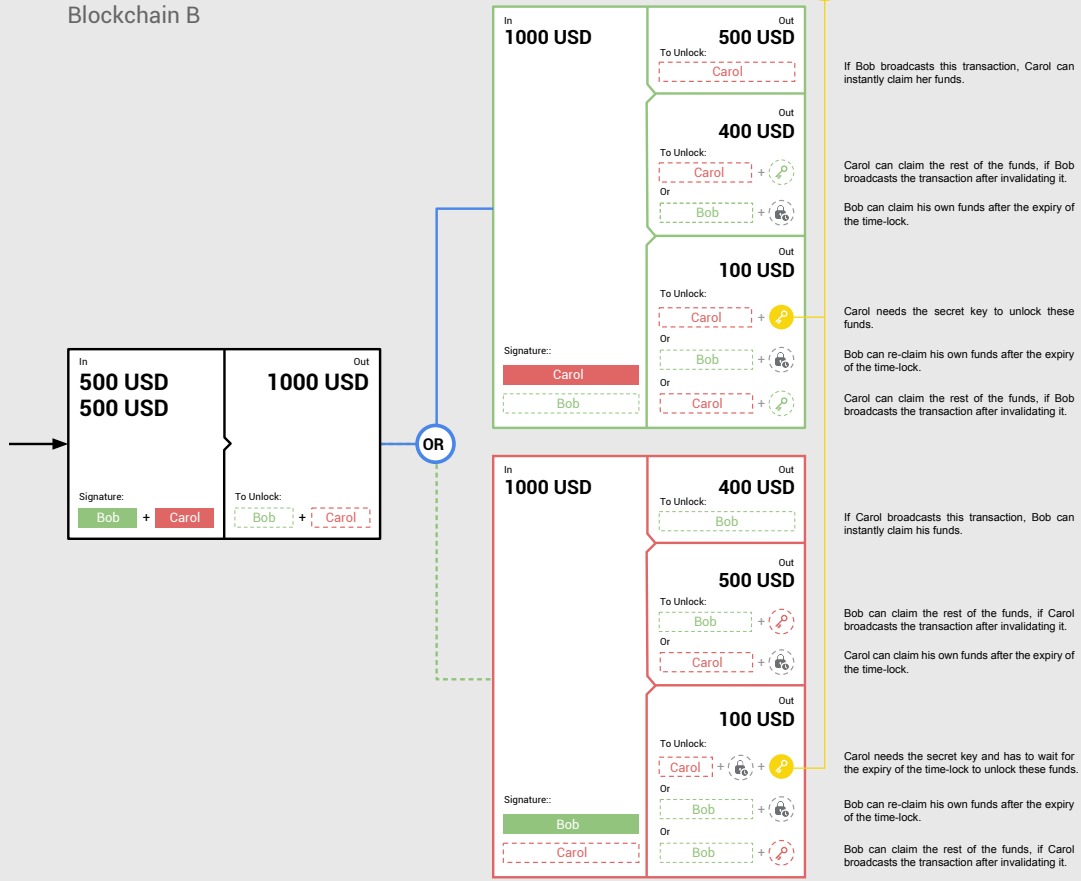

2.2.4. COMIT Routing Protocol (CRP)

A routing protocol that is private to Users, scalable in terms of network size and DOS-resistant for Liquidity Providers is an important factor. The first version of the CRP protocol is based on the BOLT 4 specifications[10] from the Lightning Network[4] extended to multi asset. This routing schema is based on the Sphinx[11] construction, and is extended with a per-hop payload, in which each LP can specify his rates for forwarding a transaction. LPs forwarding the transaction can verify its integrity, and can learn about which LP they should forward the transaction to. They cannot learn about which other LPs, besides their predecessor or successor, are part of this route, nor can they learn the length of the route and their position within it. The transaction is obfuscated at each LP, so that a network level attacker cannot associate transactions belonging to the same route.

Each User in the network learns about available LPs via a gossip protocol and constructs the route for each of his transactions transaction route, by using the public key of each intermediate LP as well as the final recipient Business or User.

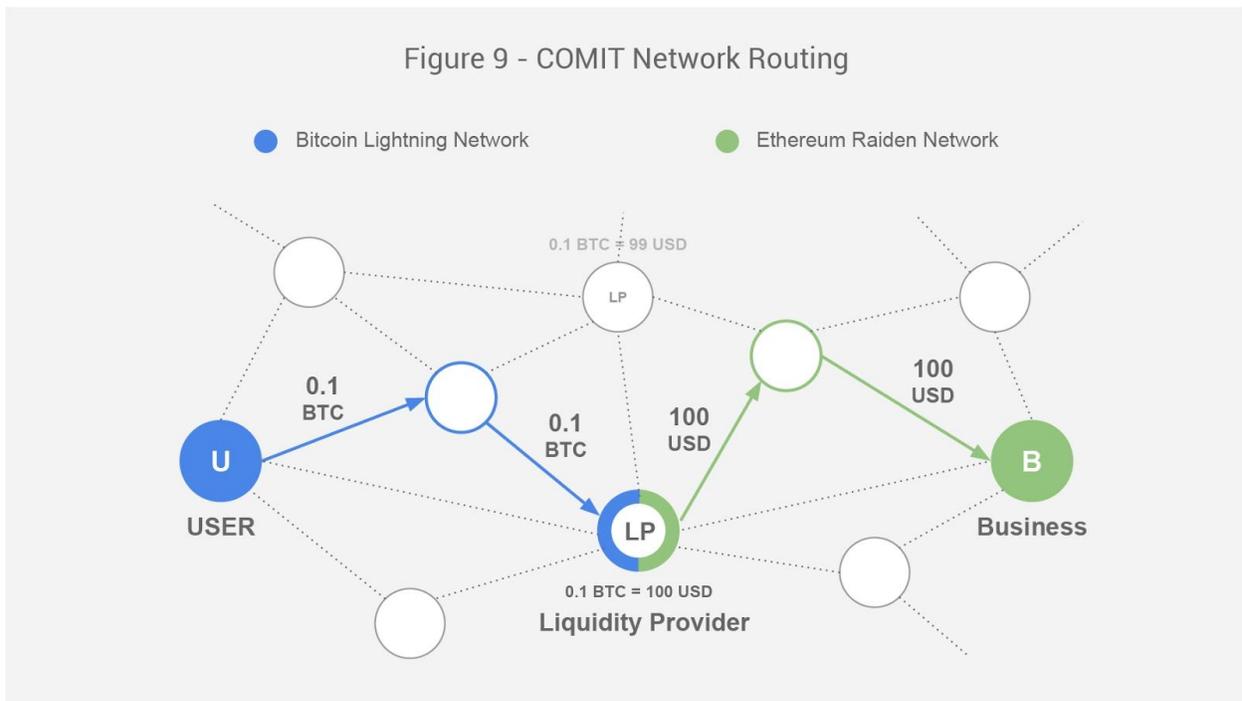

Figure 9 - COMIT Network Routing

## 2.3. How COMIT works from the perspective of Users, LPs and Businesses

COMIT consists of three different entities: We observe Users, Liquidity Providers and Businesses. Users can be on any blockchain with any asset that they wish to hold. Assets range from cryptocurrencies like Bitcoin to ICO tokens to fiat currency issued by a central bank on their own blockchain[12–14].

A Business's primary goal is to get paid. Businesses today accept one or more forms of either digital currency via the global credit card schemes, domestic payment providers and digital bank transfers. All of these payment companies are currently working on their own blockchain solution to improve efficiency and bring down transaction cost[15,16]. Our research has shown that most of them can easily be connected to COMIT to offer zero cost transactions IN and OUT of each blockchain.

Liquidity Provider bring the required liquidity to allow for seamless conversion from one asset to another.

### 2.3.1. Users (U):

A User who wants to leverage COMIT's capabilities needs to have at least one Payment Channel open to a Liquidity Provider. She can use any wallet that is compatible with COMIT to send and receive transactions on the COMIT network.

### 2.3.2. Businesses (B):

Most Businesses (B) will initially not be aware that they are connected to COMIT. They are connected to COMIT because their payment provider is choosing to upgrade his legacy infrastructure to a blockchain of his choosing. In the future we can see blockchain provider cater directly to the specific needs of businesses.

### 2.3.3. Liquidity Provider (LP):

Liquidity Provider operate on top of multiple blockchains. They provide the liquidity to convert from one blockchain asset to another. When operating between two public blockchains, they act as market makers in a decentralized marketplace. When operating with private chains, they comply with the more stringent business logic and processes that are expected to be found on such blockchains.

## 2.4. Advantages of COMIT

To summarize the advantages of COMIT, we would like to lay them out in detail for each of the three participating groups.

2.4.1. Users [U] enjoy:

- Low Cost: costs in COMIT will go towards zero as volume grows
- Instant Transaction Settlement: transactions are settled instantly no matter if multi or single asset.
- Multi Asset: any asset brought on a blockchain can be accessed through COMIT
- Access: COMIT is a superset of all accessible blockchains, giving everyone from low to high income the ability to execute transactions in any asset class.
- 100% Trust: the core infrastructure are still blockchains. Therefore neither users, liquidity providers nor businesses have to rely anyone else than the algorithm of the underlying blockchains.
- Full Control: Users retain 100% control over their assets.
- Security: The payment channels creating COMIT have in-built security mechanisms to make sure the liquidity providers can't cheat.

2.4.2. Liquidity Provider [LP]:

- Recurring Revenue Stream: Initially a pay-per-transaction and percentage model is used. At later stages a subscription to mobile providers can be envisioned.
- Low operating cost: Actual transaction costs are close to zero, because no transactions are settled onto the underlying blockchain under normal operations.
- High operating margin: LPs have high operating margins, giving them great flexibility in this business model. This does not conflict with the low cost for users as the entire transaction model is cheaper by a multitude compared to nowadays traditionally used models.
- Teamwork instead of competition: In COMIT, banks can become an integral part as LPs, thereby benefiting from any transaction executed on any blockchain.

2.4.3. Businesses [B]

- Global Reach: Just like the Internet provided massive marketing strategies on a global scale for many businesses, COMIT will expand the financial reach. This will bring new customers and more revenue.
- New markets: Today roughly 38% of the world's adult population do not have access to any form of financial service[17]. COMIT's low fees and accessibility offer huge opportunities for consumers as well as businesses especially in these underbanked markets.
- Strengthening existing markets: Especially second-world markets have seen gains in accessibility over the past year. COMIT will speed up their development even further by providing scale to otherwise financially limited areas.
- Lower financial cost: Financial cost is always a big factor for any business. Increasing the margin, increases a business's profitability and therefore its chance to succeed.
- New services possible: Just like the internet, because the speed and cost improved exponentially, made entire new business models possible, creative and entrepreneurial people will find ways to leverage COMIT's massive advantages in unforeseeable, yet highly productive ways.

# 3. Conclusion

COMIT will disrupt the financial and transaction industry just like the Internet did with media, communication and information. With COMIT we will see an even bigger push towards blockchain technologies by companies who are already in this space, further more, companies that have so far been hesitant, are now appreciating the low entry barrier. Our own focus will solely be dedicated to this area; further research and white papers will follow.

# 4. Acknowledgements

We want to thank anyone not directly mentioned in this white paper in supporting and assisting us over all this time. We also want to thank the entire blockchain community for laying the ground base over the past 8 years to make COMIT possible. A special thank you goes out to anyone involved with the Lightning Network for your groundbreaking research.